# High Spatio-Temporal-Resolution Detection of Chlorophyll Fluorescence Dynamics from a Single Chloroplast with Confocal Imaging Fluorometer

Yi-Chin Tseng[1] and Shi-Wei Chu[1,2*]

([1] Department of Physics, National Taiwan University, Taipei, Taiwan;
[2] Molecular Imaging Center, National Taiwan University;
Correspondence*: swchu@phys.ntu.edu.tw)

**Abstract**

**Background**: Chlorophyll fluorescence (CF) is a key indicator to study plant physiology or photosynthesis efficiency. Conventionally, CF is characterized by fluorometers, which only allows ensemble measurement through wide-field detection. For imaging fluorometers, the typical spatial and temporal resolutions are on the order of millimeter and second, far from enough to study cellular/sub-cellular CF dynamics. In addition, due to the lack of optical sectioning capability, conventional imaging fluorometers cannot identify CF from a single cell or even a single chloroplast.

**Result and Discussion:** Here we demonstrated a novel fluorometer based on confocal imaging, that not only provides high contrast images, but also allows CF measurement with spatiotemporal resolution as high as micrometer and millisecond. CF transient (the Kautsky curve) from a single chloroplast is successfully obtained, with both the temporal dynamics and the intensity dependences corresponding well to the ensemble measurement from conventional studies. The significance of confocal imaging fluorometer is to identify the variation among individual chloroplasts, e.g. the half-life period of the slow decay in the Kautsky curve, that is not possible to analyze with wide-field techniques. A linear relationship is found between excitation Intensity and the temporal positions of peaks/valleys in the Kautsky curve. In addition, an interesting 6-order increase in excitation intensity is found between wide-field and confocal fluorometers, whose pixel integration time and optical sectioning may account for this substantial difference.

**Conclusion:**

Confocal imaging fluorometers provide micrometer and millisecond CF characterization, opening up unprecedented possibilities toward detailed spatiotemporal analysis of CF transients and its propagation dynamics, as well as photosynthesis efficiency analysis, on the scale of organelles, in a living plant.

## Background

Chlorophyll fluorescence (CF) has been proven to be one of the most powerful and widely used techniques for plant physiologists [1-7]. Despite of its low quantum efficiency (2% to 10% of absorbed light [8]), CF detections are meaningful due to its intricate connection with numerous processes taking place during photosynthesis, such as reduction of photosystem reaction centers, non-photochemical quenching, etc. [9, 10]. It is well known that the efficiency of photosynthesis can be derived from CF dynamics, thus providing noninvasive, fast and accurate characterization for photosynthesis. It has been widely adopted to study plant physiology, including stress tolerance, nitrogen balance, carbon fixation efficiency, etc. [11]. It is not too exaggerated to say that nowadays, no investigation about photosynthetic process would be complete without CF analysis.

Conventionally, the tool of choice to study CF is a fluorometer. There are many different fluorometry techniques, such as plant efficiency analyzer (PEA) [12], pulse amplitude modulation (PAM) [13], the pump and probe (P&P) [14, 15] and the fast repetition rate (FRR) [16]. It is interesting to note that these various detection approaches are all based on the same principle, i.e. the Kautsky effect [7], or equivalent, CF transient when moving photosynthetic material from dark adaption to light environment.

Conventional imaging fluorometers (e.g., PAM and P&P fluorometers) are based on wide-field detection, and are routinely adopted to study ensemble of CF transients from a large area of a leaf, significantly limiting its spatiotemporal resolution. For example, to study stress propagation in a plant leaf [17], current imaging fluorometers only provide spatial resolution on the order of sub-millimeter, with temporal resolution on the order of second. To unravel the more detailed propagation dynamics, the required spatial resolution should be at least on single cell or sub-cellular level, while the temporal resolution should be enhanced to millisecond scale.

One additional drawback of the conventional imaging fluorometers is lack of optical section capability due to their wide-field nature, and thus prevents study of CF transient on a single cell or even a single chloroplast level. In this work, we introduce a novel concept of confocal imaging fluorometer, which is the combination of confocal microscopy and CF transient detection, where the former provides optical sectioning with exceptionally high axial contrast. The technique not only detects CF signals with microsecond temporal resolution, but also attains micrometer spatial resolution in all three dimensions. We have successfully reproduced the CF transient (Kaustky curve) within a single chloroplast. We found that the CF transients of a group of palisade cells and the ensemble of single chloroplasts are similar to each other, and both correspond well to the result of conventional imaging fluorometers, showing the reliability of our result. Nevertheless, the CF transient of individual chloroplast can be substantially different, manifesting the value of the unusual capability to study plant cell organelles. Furthermore, we found that the shape of transients is highly intensity-dependent, which was also shown in the previous research [18]. Finally, short integration time and optical section characteristic of confocal image fluorometer make a significant difference of illumination intensity comparing to that of conventional fluorometers. Given CF transient from a single chloroplast, it is possible to investigate degree of influence from external or internal plant-stress with scale of organelle, and confocal imaging fluorometer has paved the way for this high spatio-temporal resolution CF detection.

## Principle

### *Basic Concept of Confocal Imaging Fluorometer*

The optical principle of confocal imaging fluorometer is basically the same as confocal laser-scanning microscopy [19], which is an optical imaging technique for increasing contrast and resolution. The essential components of a confocal imaging fluorometer is shown in Figure 1, including a laser system, a dichroic mirror, a scanning mirror system, an objective lens, a pinhole and a photomultiplier tube (PMT).

The laser system in a confocal imaging fluorometer provides strong and monochromatic illumination, whose wavelength can be selected to meet sample request. The laser beam is sent to the objective after the scanning mirror system to achieve two-dimensional raster scanning at the focal plane. The backward fluorescence signal is collected by the same objective, de-scanned through the scanning mirrors, and separated from residual laser by the dichroic mirror. The fluorescence signal then is focused onto the pinhole, which is placed at the conjugate plane of objective focus, to achieve optical sectioning by excluding out-of-focus signals. One or more PMTs are placed behind the pinhole to collect the in-focus fluorescence signals, which are reconstructed into images by synchronization with the scanning mirrors [19].

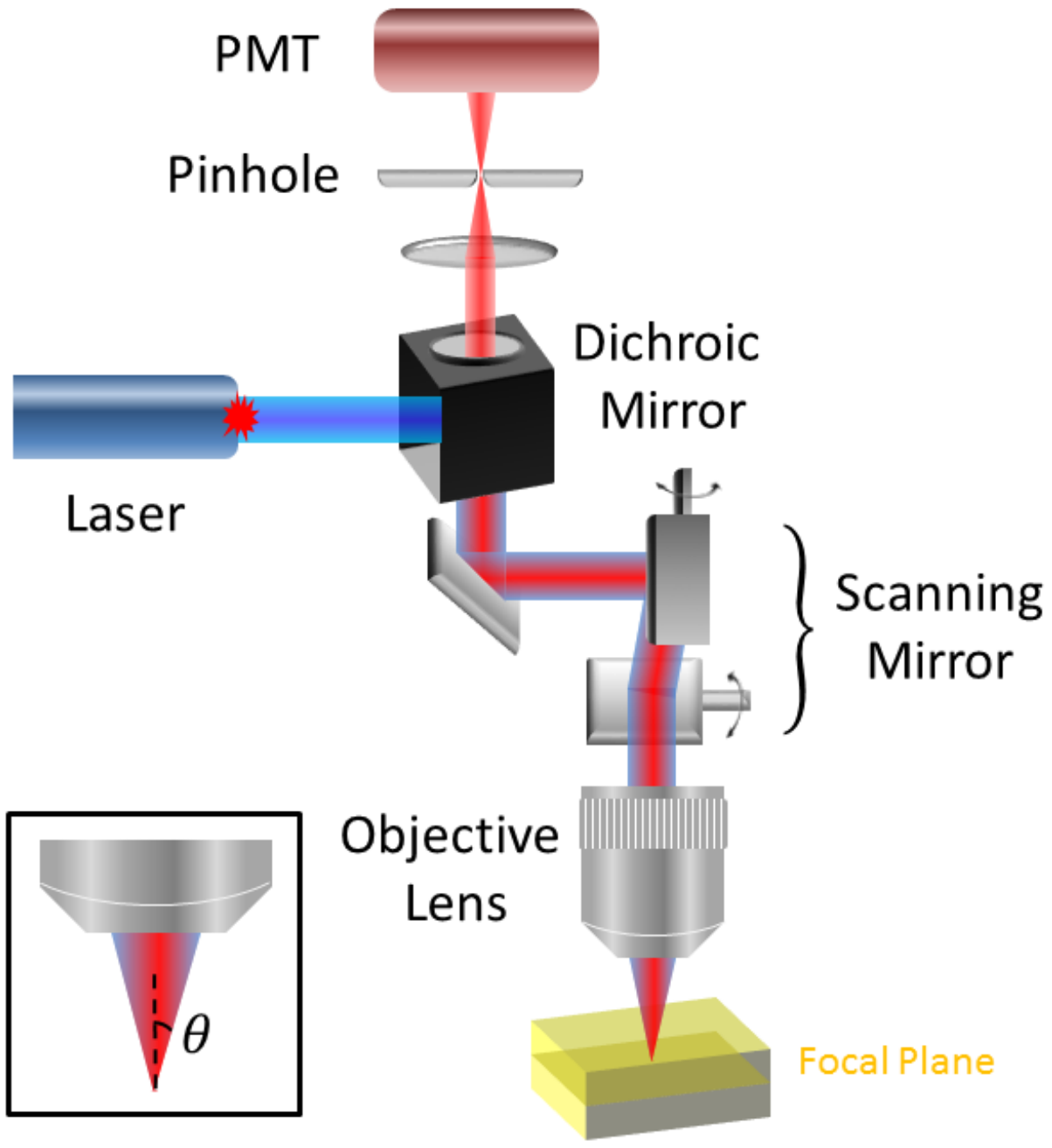

**Figure 1** Principle and basic components of a confocal imaging fluorometer. Laser beam is reflected by a dichroic mirror and goes through a set of scanning mirrors, then focused by an objective lens onto the specimen. Fluorescence signals is epi-collected in the same path, and filtered out by the dichroic mirror. A confocal pinhole is used to allow only fluorescence emitted from the focal plane being detected by the PMT.

In general, a confocal imaging system is capable of collecting signal with a well-defined optical section on the order of 1 μm [20]. This high axial resolution makes confocal system an invaluable tool to observe single cell or sub-cellular organelles [21-23]

The objective lens is characterized by magnification and numerical aperture (NA). To enable large field-of-view observation, low magnification objectives are typically required. However, please note that resolution is determined by NA, which can be independent from magnification. NA describes the light acceptance cone of an objective lens and hence light gathering ability and resolution. The definition of NA is:

$$NA \equiv n \times \sin\theta \tag{1}$$

In the definition of NA, n is the index of refraction of the immersion medium, while θ is the half-angle of the maximum light acceptance cone. Both lateral (xy-direction) and axial (z-direction) resolutions for fluorescence imaging mode are defined by NA and the wavelength (λ) [24].

$$r_{lateral} = \frac{0.43 \times \lambda}{NA} \quad (2)$$

$$r_{axial} = \frac{0.67 \times \lambda}{n - \sqrt{n^2 - NA^2}} \quad (3)$$

### *Kautsky Effect*

Kautsky effect, first discovered in 1931, describes the dynamics of CF when dark-adapted photosynthetic chlorophyll suddenly exposes to continuous light illumination [25]. After initial light absorption, chlorophyll becomes excited and soon releases its energy into one of the three internal decay pathways, including photosynthesis (photochemical quenching, qP), heat (non-photochemical quenching, NPQ) and light emission (CF). Owing to energy conservation, the sum of quantum efficiencies for these three pathways should be unity. Therefore, the yield of CF is strongly related to the efficiency of both qP and NPQ [26].

To be more specific, when transferring a photosynthetic material from dark adaption into light illumination, CF yield typically exhibits a fast rising phase (within 1 second) and a slow decay phase (few minute duration), as shown by the green curve in Figure 2. The fast rising phase is labeled as OP, where O is for origin, and P is the peak [18]. It is mainly caused by the reduction of qP; that is, depletion of electron acceptors, quinine (Qa) in the electron transport chain [27]. The slow decay phase is labeled as PSMT, where S stands for semisteady state, M for a local maximum, and T for a terminal steady state level [18]. One very interesting phenomenon is the shape of this decay phase depends strongly on illumination intensity. At low intensity (32 $\mu mol/m^2/s$), the Kaustky curve is the green one. When the intensity grows one order larger, the amplitude of SM-rise in the transient is smaller, as shown by the red curve. At one more order higher intensity, the SM section disappears completely, leaving an exponential decay in the PT section, as shown by the blue curve. Such intensity-dependent curve transition is known to be the result of photosynthetic state transition, on which more detailed discussion can be found in the references [1, 10, 13, 28-30].

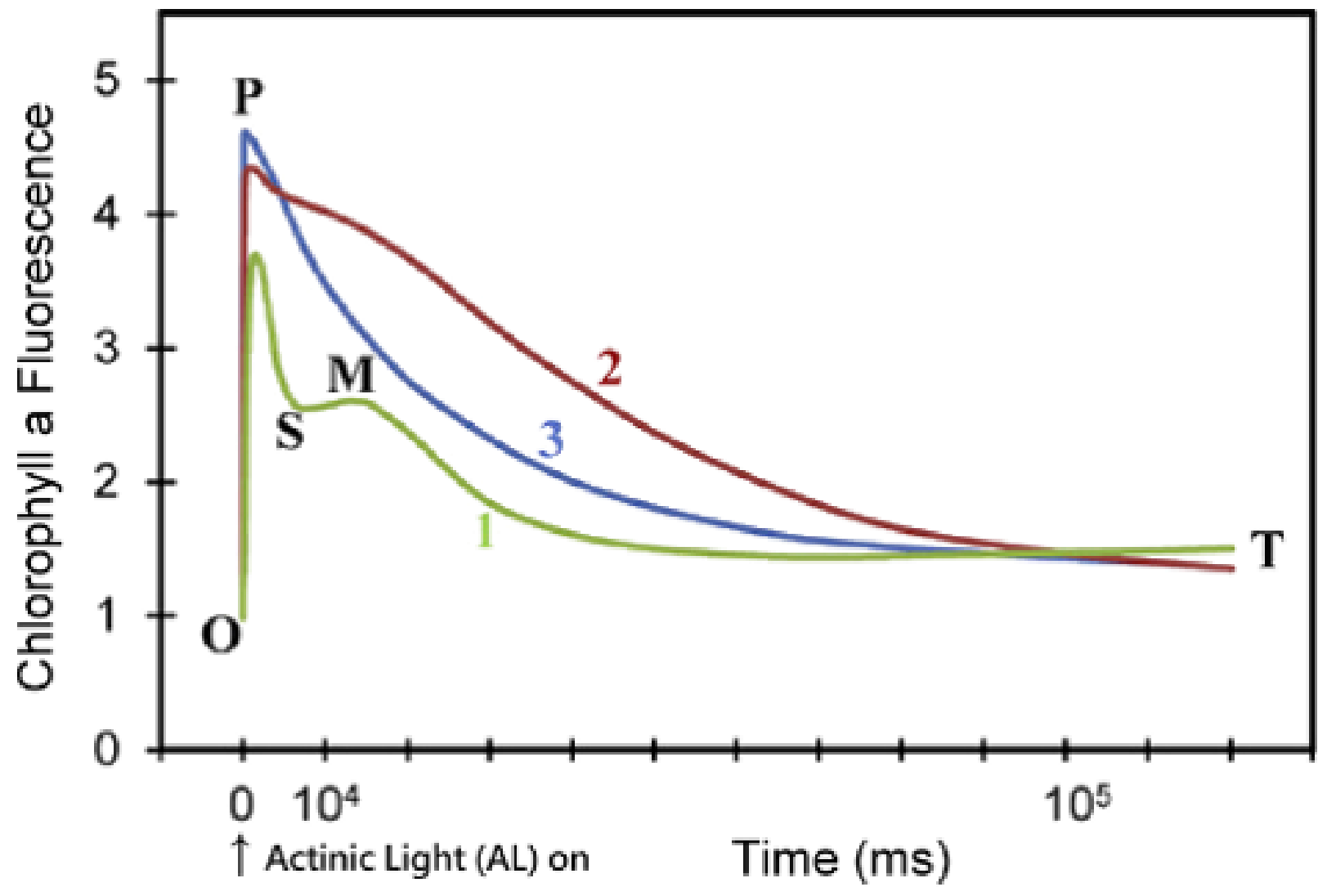

**Figure 2** The Kautsky effect, showing the CF transient as well as its intensity dependence. Wavelength of excitation: 650nm. Excitation light intensity for curves labeled 1, 2 and 3 was 32, 320 and 3200 μmol/m$^2$/s, respectively. For definition of OPSMT, O is the origin, P is the peak, S stands for semi-steady state, M for a local maximum, and T for a terminal steady state level. (Modified figure from [1], with copyright permission)

## Material and Method

### *Plant Material*

*Brugmansia suaveolens* (solanaceae), also known as Angel's Trumpet, is a woody plant usually 3 m to 4 m in height with pendulous flowers and furry leaves distributed widely in Taiwan, especially in wet areas. Being interested in spatial temporal dynamics of CF, we selected *B.suaveolens* as our target material since the CF of its cousin *Datura wrightii*, also known as Devil's Trumpet, had been studied in depth [17]. *B.suaveolens* leaves were collected from the Botanical Garden of National Taiwan University, Taipei, Taiwan (25°1' N, 121°31' E, 9 m a.s.l.). All sample leaves are picked as fully expanded leaves that had neither experienced detectable physical damage nor herbivory. In order to minimize the sampling error, 3 leaves are chosen within plants that grow in similar micro-climate. Furthermore, all the measurements are completed no longer than two hours after disleaving. Fresh leaves are sealed in slide glass (76 × 26 mm), and slide samples are dark-adapted under constant temperature and constant humidity dark environment (20 ºC, 70 %RH) for 20 min.

## *Experimental Setup*

A confocal microscope (Leica TCS SP5) in the Molecular Imaging Center of National Taiwan University was adopted. CF was excited by a HeNe laser, whose wavelength (633 nm) is the same as that used in popular conventional fluorometers, such as LI-6400 from LI-COR. A relatively low-NA objective (HC PL Apo 10x/0.4 CS) was selected to allow not only large field of view over a few millimeters, but also resolution much better than a single chloroplast. From Eq. (2) and (3), The lateral and axial resolutions are about 1 μm and 5 μm, respectively.

The initial step is to bring the sample to focus by weak excitation (~1 kW /cm$^2$), and then the leaf is left in dark again for 5 minutes. To observe the Kautsky effect, the 633-nm laser was focused on the sample, and the fluorescence emission was recorded in the spectral range of 670 to 690 nm. The intensity-dependent CF transient curves were obtained by taking time-lapsed images while varying the 633-nm excitation intensity from 1 kW/cm$^2$ to 50 kW/cm$^2$, at different sample regions. With different number of total pixels, the temporal resolution of the CF transient varies from 10 milliseconds (16x16 pixels) to about 200 milliseconds (256x256 pixels). No significant photobleaching of CF is expected at this intensity range [31].

# Results

## *Fluorescence dynamics from a single chloroplast*

Conventional fluorometers observe CF dynamics over a large area on a leaf, and here we demonstrate that our confocal imaging fluorometer allows us to obtain CF transients from a precisely chosen cells or even a single chloroplast. Figure 3(a) shows the confocal images of the leaf sample. Figure 3(a1) is the large-area view, showing the distribution of vascular bundles. (a2) gives a zoom-in view of a group of palisade cells, showing clear distribution of chloroplasts in each cell. By further zooming in, the field of view is focused onto a single chloroplast, as given in (a3), showing the distribution of chlorophyll density inside the organelle [32].

Fig. 3(b) presents the CF transients from a group of palisade cells (b1) and a single chloroplast (b2). Both curves are normalized with its own maximal fluorescence intensity. The characteristic slow decay of Kautsky curve are obvious in both curves, but the fast rising phase is only visible in (b1), because (b2) is noisier due to less pixels involved. By fitting the curves with exponential decay function, the time constants are found to be 28 ± 0.5 s for (b1) and 25 ± 1.8 s for (b2). The numbers correspond well to the large-area measurement with conventional fluorometers [33], manifesting the reliability of the confocal fluorometer. However, this result also demonstrates that the CF transients can be slightly different among individual chloroplasts and cells, as shown in Fig. 4. Such detailed characterization is not possible with conventional fluorometers, which only deliver the ensemble response from a lot of cells.

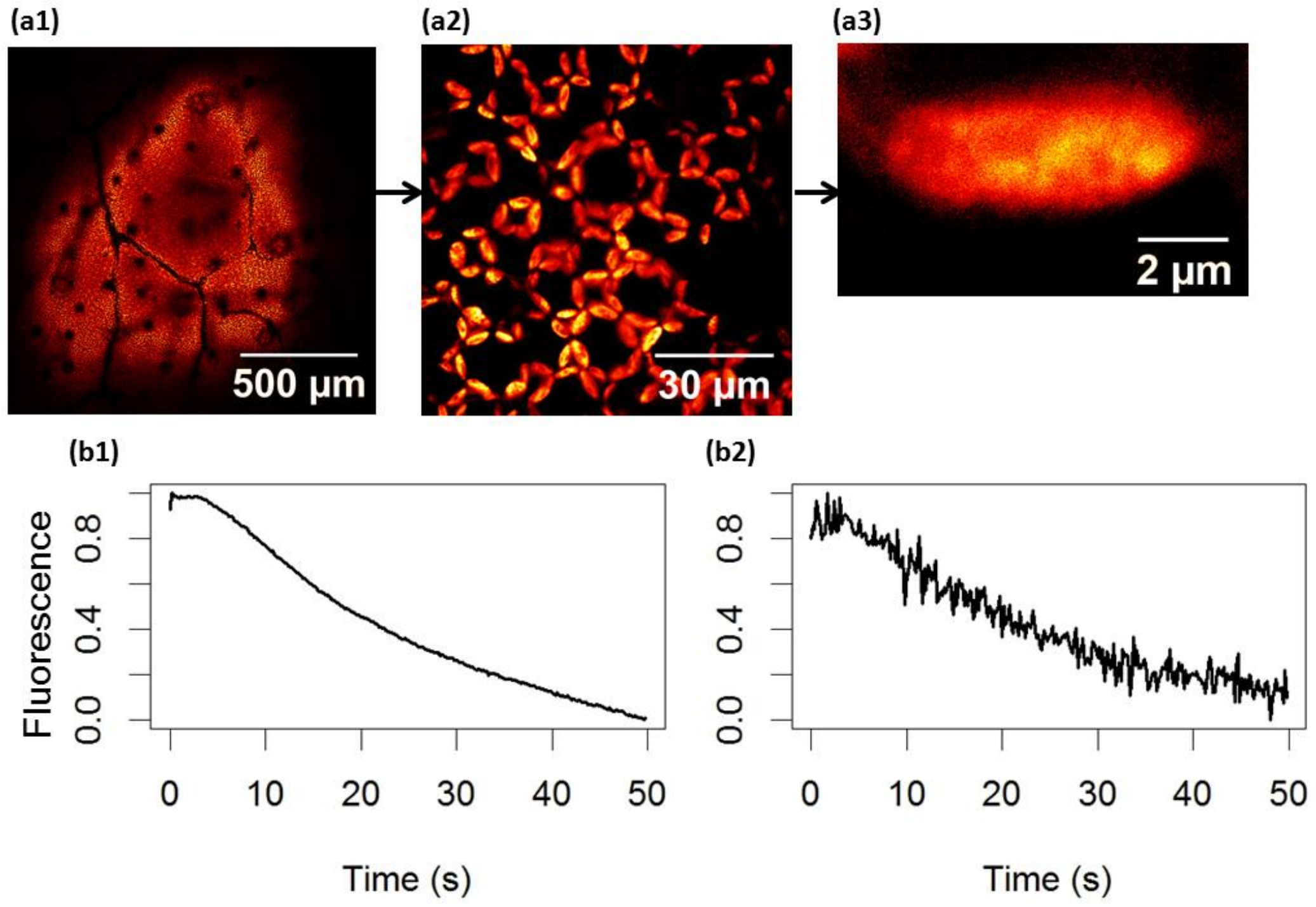

**Figure 3** Confocal images and CF transients on different spatial scales inside a living leaf. A 633-nm laser, with 40 $kW/cm^2$ intensity, is adopted for 50 s continuous confocal imaging. The sample leaf was kept in darkness for ~20 min before imaging. (a1) is the image over a large area of the leaf, (a2) is zoomed in to show a group of palisade cells, and (a3) is further zoomed in to focus onto a single chloroplast. (b1) and (b2) show normalized CF transient from a group of palisade cells and a single chloroplast, respectively. (b2) is noisier since less pixels are involved.

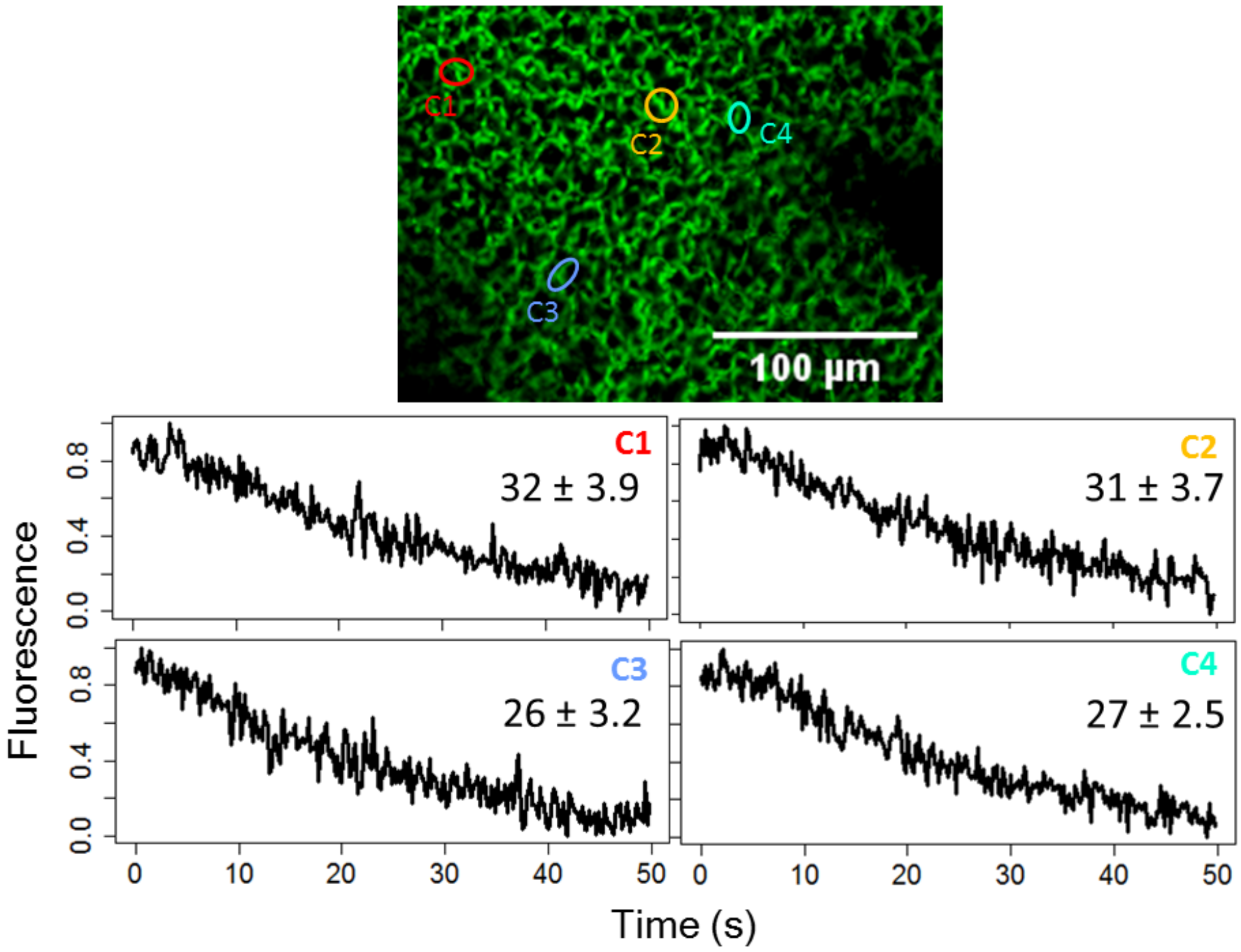

**Figure 4** Fluorescence transients for individual chloroplasts within a living leaf, which is confocally imaged at 40 kW/cm$^2$ with a HeNe laser (633nm).

## *Intensity dependent fluorescence transient*

As we have mentioned in Figure 2 of the Principle section, it is well known that the Kautsky curve changes with intensity. Figure 5 shows the intensity-dependent Kautsky curves from a single chloroplast (gray lines) and from a group of cells (colored lines), obtained by the confocal fluorometer. Fig. 5(a) is acquired with low laser intensity (3 kW/cm$^2$), and a temporal variation similar to curve 1 of Fig. 2 is found, i.e. a complete OPSMT curve. The CF intensity rises to its first peak within 1 s, quickly decreasing to a local minimum (i.e., PS-decrease), rising again to a second peak (i.e., SM-rise) then slowly falling as exponential decay. At slightly higher intensity (10 kW/cm$^2$), a temporal variation similar to curve 2 of Fig. 2 is observed. The PS-fall and SM-rise still exist, but become much smaller, while the position of P, S, and M appear earlier in the curve. At high intensity (55 kW/cm$^2$), the SM part disappears completely, leaving a single exponential decay (the PT section), similar to curve 3 of Fig. 2. This result matches very well to the conventional wide-field fluorometer [1, 13, 29], but with much higher spatio-temporal resolution, manifesting again the reliability and usefulness of the confocal technique.

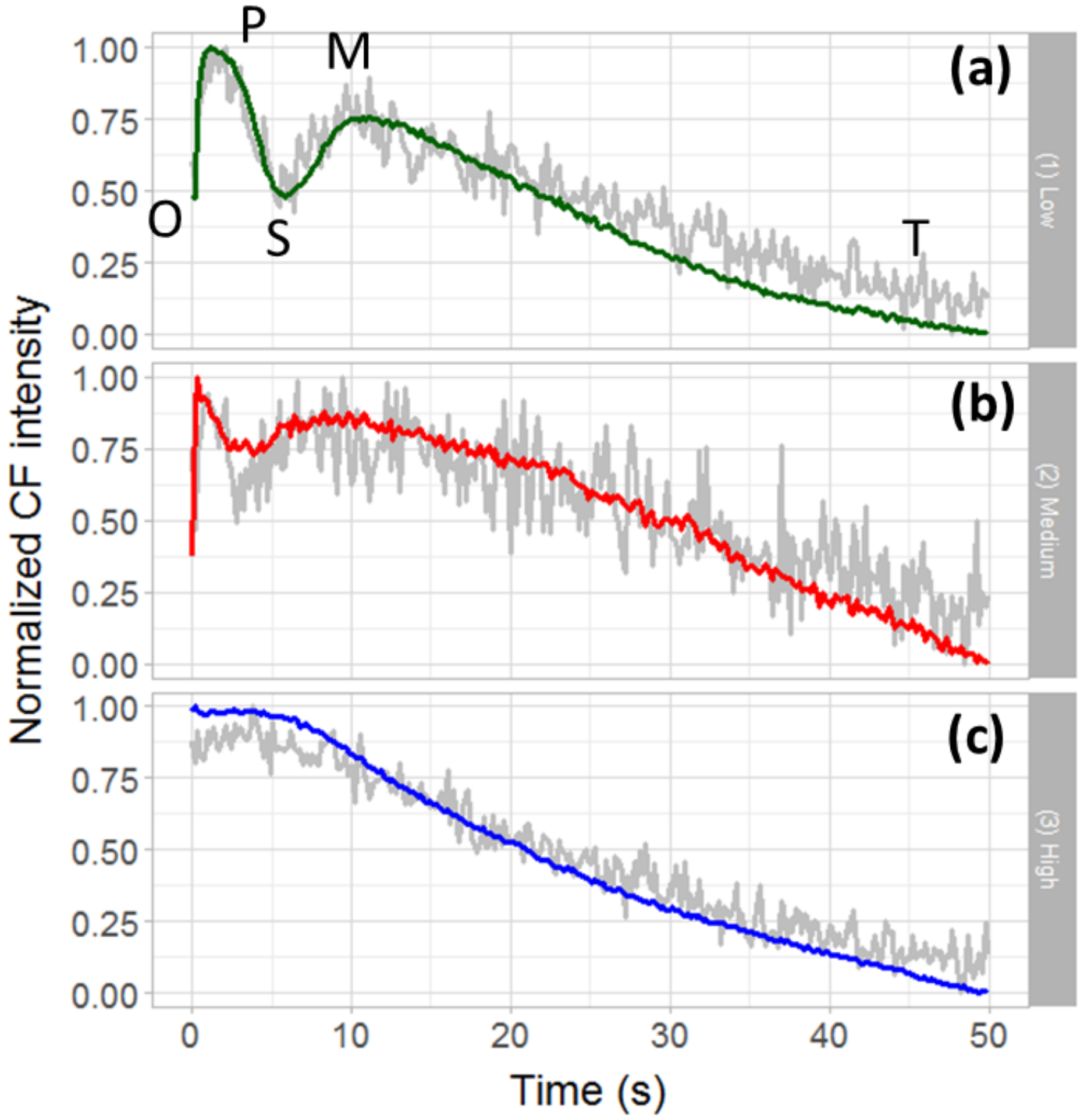

**Figure 5** CF transients of a single chloroplast (grey) and average among a group of cells (colored) under excitation intensity at (a) 3, (b) 10, and (c) 55 kW/cm$^2$, respectively, showing clearly the intensity-dependent Kautsky curves.

An interesting observation in Fig. 5 is not only the curve shape is intensity-dependent, but also the positions of local maxima and minimum P, S, M points are also strongly dependent on excitation intensity. Fig. 6(a) shows the detailed curve variation relative to intensity, in the range of 3- 55 kW/cm$^2$, and the corresponding temporal position of local maximum of induced transients, i.e., point M, is given in Fig. 6(b). Surprisingly, an almost perfect linear trend is observed. Similar linear results are found for the semisteady state point S in Fig. 6(c), and for the peak point P in Fig. 6(d). Due to the limitation of temporal resolution (200 ms for 256x256 pixels), S and P points are analyzed with intensity range 3- 40 kW/cm$^2$ and 3- 20 kW/cm$^2$, respectively. The linear trends indicate that the state transition rate increases with higher excitation intensity. The underlying mechanism relies more investigation in the future.

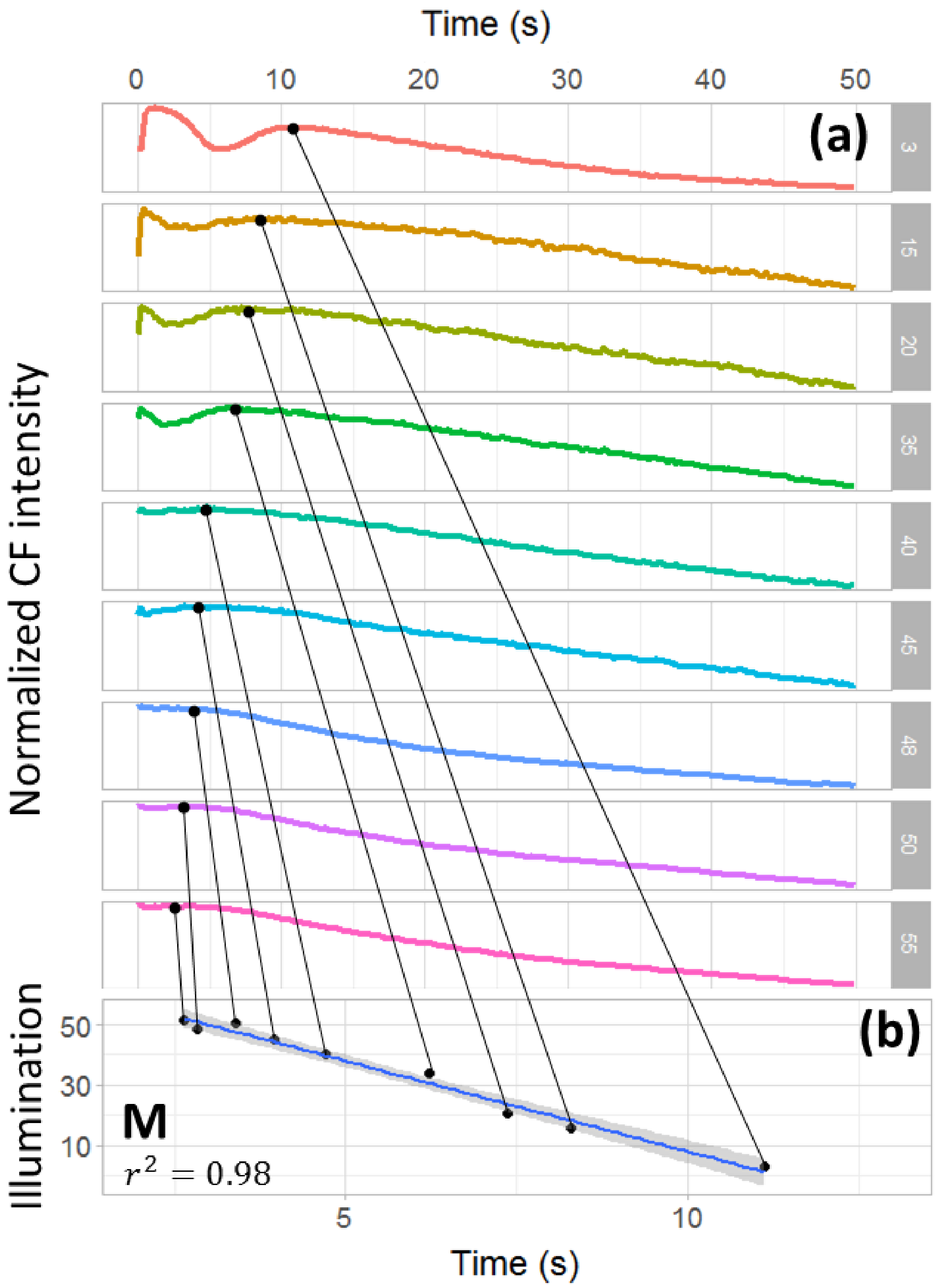

**Figure 6** (a) Detailed Kautsky curve variation in the intensity range of 3 – 55 kW/cm$^2$ The temporal positions of (b) the local maximum (M), (c) semisteady state (S), and (d) peak (P), all change linearly with excitation intensity. The Grey area represents 95% confidence region.

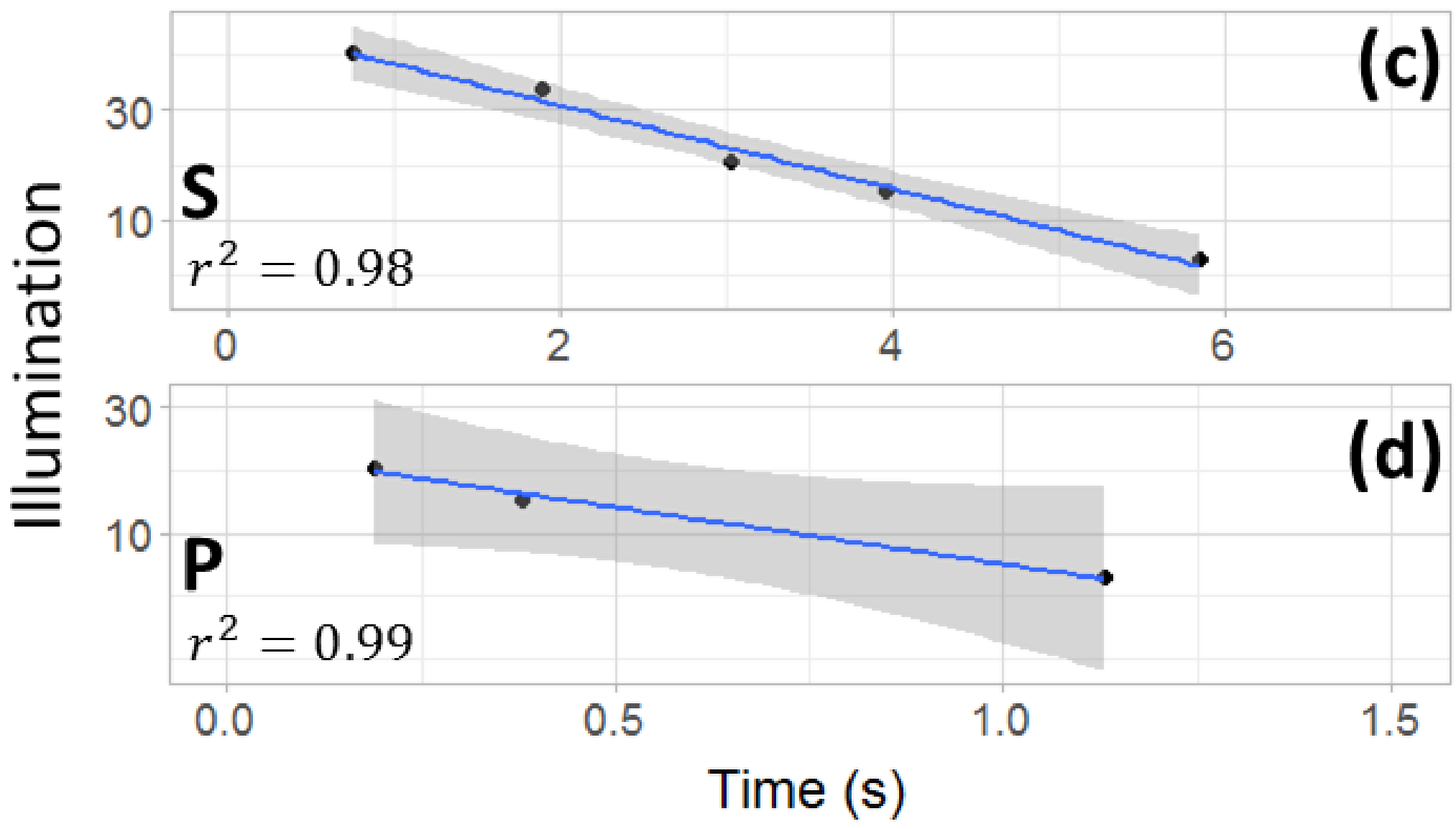

**Figure 6(c)~(d)** An analysis of semisteady state (S) and peak (P) in induced transients with different illumination intensity. The laser intensity was applied from 3- 40 kW/cm$^2$ for S analysis while 3- 40 kW/cm$^2$ for P analysis. The responding time ranges from 1- 6 seconds for S state; 0.3- 1.2 second for P state. Upper panel: linear regression of illuminating intensity and time for semisteady state in transients (S). Lower panel: linear regression of illuminating intensity and time for peak in transients (P). Grey area represents 95% confidence region.

## Discussion

We have successfully obtained the Kautsky curve, as well as its intensity dependence, with the confocal imaging fluorometer. Comparing to conventional wide-field imaging fluorometers, the confocal technique allows much better spatial confinement due to optical sectioning capability, and thus observation from a single chloroplast becomes possible. It's very interesting to notice that the Kautsky curves acquired by the confocal fluorometer are similar to those acquired by the wide-field fluorometers. In terms of the half-life period of the slow decay, the ensemble results of the confocal fluorometer agree very well to the wide-field ones, but only the confocal fluorometer is capable to identify the difference among individual plant cells or chloroplasts.

In terms of the temporal resolution performance, the confocal and wide-field fluorometers should be similar in terms of a single pixel detection, which takes about 1 – 10 μs in both cases. As we mentioned in [17], the wide-field fluorometer takes about 1 second to record one image. Nevertheless, the advantage of the confocal scheme is the freedom to select number of pixels, as well as the position of these pixels, significantly enhancing the temporal responses. By using more advanced scanning approaches, such as random-access microscopy [34], high-speed CF detection among distant chloroplasts is possible. In addition, by adopting a multi-focus scanning approach, such as being demonstrated by spinning disk confocal microscopy in 2009 [35], the frame rate of confocal fluorometer can be significantly improved. Please note that when using the spinning disk technique, the illumination intensity has to be calculated by the focus spot size, not by the imaging area, so the intensity description in [35] should be multiplied by at least $10^3$.

Another important aspect to notice is that the illumination intensity of the confocal fluorometer is much higher than that of the wide-field fluorometers. As shown in Fig. 5 and Fig. 6, to eliminate the semi-steady state S in the CF transient, about 55 $kW/cm^2$ is required for the confocal fluorometer. However, in the case of wide-field fluorometer, as shown in the example of Fig. 2 [1], to eliminate S, 3200 $\mu mol/m^2/s$ is required. Considering the wavelength to be 650 nm in [1], the photon energy is $1240/650 = 1.9$ eV $= 3 \times 10^{-19}$ J. Therefore, the intensity unit ($\mu mol/m^2/s$) is equivalent to $[10^{-6} \times 6 \times 10^{23}$ (# of photons)$] \times [3 \times 10^{-19}$ (J/photon)$] / 10^4$ $cm^2$ / s $= 18 \times 10^{-9}$ $kW/cm^2$. As a result, in the wide-field fluorometer, the required illumination intensity is $3200 \times 18 \times 10^{-9}$ $kW/cm^2 = 5.76 \times 10^{-5}$ $kW/cm^2$, six orders of magnitude smaller than that in the confocal one.

To explain this 6-order intensity difference, optical sectioning and illumination time of the confocal imaging fluorometer have to be considered. In a conventional fluorometer (wide-field detection), CF signals are emitted throughout the whole leaf in the axial direction, so the depth of field (i.e. signal collection depth) is equivalent to the thickness of a leaf, which is usually 100-1000 μm. On the other hand, for a confocal fluorometer, a pinhole is inserted before the detector to reject most out-of-focus fluorescence, and thus the total signal strength is significantly reduced. The typical depth of field in a confocal fluorometer is about 1-10 μm, which is two orders less than that of the wide-field one. Hence, the signal strength of the confocal fluorometer is expected to be two orders weaker than the wide-field counterpart.

In terms of the illumination time, in a conventional wide-field imaging fluorometer, the whole leaf sample is illuminated continuously, so the illumination time for each pixel is the same as the frame acquisition time. On the other hand, a small laser focus scans across the sample in the confocal scheme, making the illumination time for each pixel much shorter than the frame time. For example, in the case of Fig. 4(a2), one frame takes about 1 second, and the frame is composed of $256 \times 256$ pixels, so the illuminating time for each pixel (1 pixel is roughly 1 $\mu m^2$ in this case) of the confocal imaging fluorometer is about four orders shorter than that of conventional wide-field imaging fluorometer.

Combining the above two reasons, it is reasonable that the illumination intensity in the confocal imaging fluorometer needs to be much higher than that in the wide-field fluorometer to achieve similar CF signal strength, as well as the Kautsky curves. The latter is somewhat surprising since it indicates that the physiological response of the chlorophyll remains the same with such high-intensity, yet short-period, illumination. One possible reason is that there is a slow reaction during photosynthesis and CF generation, so the chlorophyll only “feels” the average intensity, not the instantaneous intensity. By looking into the electron transport chains in the photosystem, the bottleneck reaction might be the reduction of plastoquinone (PQ), which has a relatively slow reaction rate (100 molChl $mmol^{-1}$ $s^{-1}$) [36]. Further studies are necessary to identify the underlying photochemical mechanism.

## Conclusion

In this work, we demonstrated a novel confocal imaging fluorometer that can provide high spatiotemporal characterization of CF inside a living leaf. The three-dimensional spatial resolution is on the order of micrometer, and the temporal resolution reaches tens of milliseconds, allowing us to study CF transient, i.e. the Kautsky effect, from even a single chloroplast. Although the ensemble behavior of CF transient, as well as the intensity-dependent Kautsky curves, agree well with the results of conventional wide-field fluorometers, confocal imaging fluorometer provides valuable information toward the difference of CF decay rate among individual chloroplasts. The features of optical sectioning and laser focus scanning in the confocal fluorometer result in much higher illumination intensity compared to conventional techniques, while maintaining normal cellular physiological responses. Our work not only opens up new possibilities to study CF dynamics on the level of organelles, but also is promising to unravel more spatial/temporal details in the associated photosynthetic processes.

### Competing interests

The authors declare that they have no competing interests.

### Author’s contribution

YCT designed the experiment, carried out signal analysis, and wrote most of the manuscript. SWC envisioned the idea, provided the experimental hardware, and helped to polish the manuscript.

### Acknowledge

The authors appreciate the inspirational discussion with Prof. Govindjee from University of Illinois at Urbana-Champaign. This work is supported by the Molecular Imaging Center of NTU (105R8916, 105R7732), and by the Ministry of Science and

Technology, Taiwan, under grant MOST-102-2112-M-002-018-MY3 and MOST-105-2628-M-002-010-MY4. SWC acknowledge the generous support from the Foundation for the Advancement of Outstanding Scholarship.

**Author details**